\documentclass[aip,jcp,graphicx,reprint]{revtex4-1}
\usepackage{graphicx}
\usepackage{amsmath}
\usepackage{amssymb}
\usepackage{hyperref}

\begin{document}

\title{Accurate numerical integration of an electron exchange hole with a screened Coulomb interaction}

\author{Jonathan E. Moussa}
\email{godotalgorithm@gmail.com}

\author{Peter A. Schultz}

\affiliation{Sandia National Laboratories, Albuquerque, NM 87185, USA}

\date{\today}

\begin{abstract} \centering \begin{minipage}{0.79\textwidth}
The numerical implementation of an exchange-correlation functional is not always an accurate reproduction of its theoretical specification.
For example, density functionals for exchange and correlation can be defined by an exchange or correlation hole function
 that is integrated with the Coulomb interaction to form an energy.
This construction can be used to modify a density functional for use with any electron-electron interaction.
Its most prominent use is in the Heyd-Scuseria-Ernzerhof (HSE) functional
 that generalizes the Perdew-Burke-Ernzerhof (PBE) model of exchange to a screened Coulomb interaction
 with an error function form.
However, we find non-negligible numerical errors in the standard implementation of the HSE exchange hole integration.
We formulate and implement a new method for evaluating the exchange hole integration
 that is simple, accurate, and efficient.
Its numerical errors are bounded and minimized by applying basic elements of approximation theory.
\end{minipage} \end{abstract}
\maketitle

\section{Introduction}

The Kohn-Sham density functional theory (DFT) of electrons is specified by
 density functionals that model the exchange and correlation energy of electrons.
These functionals are presently limited by inaccuracies in reference data,
 insufficient physical constraints, and simplification to convenient but approximate mathematical forms.
In most cases, these functionals have an explicit analytic form and their numerical evaluation is straightforward.
However, increasing the accuracy of functionals inevitably requires more complicated forms.
More complex functionals require improved numerical methods to evaluate them both accurately and efficiently.

In this paper, we reassess the numerical evaluation of the semilocal exchange energy component
 of the Heyd-Scuseria-Ernzerhof (HSE) functional \cite{HSE03},
 which cannot be expressed exactly in a closed analytic form.
Its evaluation requires additional numerical approximations and is subject to numerical errors.
To date, it has had three distinct implementations \cite{HSE03,HSE04,HSE09}.
They differ primarily in their numerical integration of an electron exchange hole with a screened Coulomb interaction.
They also embody different regularizations of the underlying exchange hole model \cite{PBE_hole},
 which breaks down in the limit of large electron density gradients.
We reevaluate these details and present a new method for approximating the integral that is simple
 and free of discontinuities and singularities.
Our proposed reformulation is accurate to single precision ($\approx 10^{-7}$) to reduce numerical errors
 far below the physical uncertainties of the HSE functional and limit the inevitable growth of errors in functional derivatives.

It had been shown that the basic physics contained in the HSE exchange hole
 can be captured in more efficiently computable exchange hole models \cite{new_xhole}.
This is an example of using physical modeling to resolve the difficulty of numerically evaluating the HSE functional.
However, those results are not numerically identical to the HSE functional and thus these constitute a distinct density functional.
We consider an alternate approach: directly confront
 the numerical evaluation of the existing HSE functional as an application of approximation theory.
This approach makes use of several basic elements of approximation theory that include nonlinear minimax approximation,
 special function evaluation, and error analysis.
Our result serves as a definitive numerical implementation of the HSE semilocal exchange functional
 and demonstrates how increased sophistication in numerical analysis expands 
 the set of numerically tractable models that might be used in the construction of future density functionals.

The paper is organized as follows.
First, we review the HSE functional.
Second, we define a new regularization to ensure that its value and derivatives are well-defined.
Third, we derive a new scheme for numerically integrating the exchange hole
 based on an approximation of the complementary error function with Gaussians.
Finally, we assess the accuracy of our new implementation and compare
 with previous implementations for functional values and first derivatives.

\section{Summary of the HSE functional}

For clarity, we summarize the details of the HSE functional that are relevant to discussions of the exchange hole integration.
The HSE correlation energy is identical to the Perdew-Burke-Ernzerhof (PBE) functional \cite{PBE}
 and is not of interest here.
The HSE exchange energy can be written as
\begin{align}\label{exchange_energy}
 E_{\mathrm{x}}^{\mathrm{HSE}} &= a E_{\mathrm{x}}^{\mathrm{HF,SR}}(\omega)
  + (1-a) E_{\mathrm{x}}^{\mathrm{PBE,SR}}(\omega) \\ & \ \ \ + E_{\mathrm{x}}^{\mathrm{PBE,LR}}(\omega). \notag
\end{align}
The terms of this expression are defined by a splitting of the Coulomb interaction with error functions into
 a short-range ``screened'' interaction and a long-range tail,
\begin{equation}
 \frac{1}{r} = \frac{\mathrm{erfc}(\omega r)}{r} + \frac{\mathrm{erf}(\omega r)}{r}.
\end{equation}
The first term of Eq. (\ref{exchange_energy}) is the Fock exchange energy computed with the short-ranged screened Coulomb interaction,
\begin{align}
 E_{\mathrm{x}}^{\mathrm{HF,SR}}(\omega) &= \\
 -\frac{1}{2}\sum_{\sigma,\sigma'} & \int d\mathbf{r} d\mathbf{r}' 
  \frac{\mathrm{erfc}(\omega |\mathbf{r}-\mathbf{r}'|)}{|\mathbf{r}-\mathbf{r}'|} |\rho_{\sigma,\sigma'}(\mathbf{r},\mathbf{r}')|^2, \notag
\end{align}
 which requires access to the spinful Kohn-Sham density matrix, $\rho_{\sigma,\sigma'}(\mathbf{r},\mathbf{r}')$.
The remaining terms replace the density matrix with an exchange hole model \cite{PBE_hole}, $J(\mathbf{r},|\mathbf{r}-\mathbf{r}'|)$,
 that is averaged over angle and assumes no mixing of electron spins,
\begin{equation}
 |\rho_{\sigma,\sigma'}(\mathbf{r},\mathbf{r}')|^2 \approx
  - 2 \delta_{\sigma,\sigma'} \rho_\sigma(\mathbf{r})^2 J(\mathbf{r},|\mathbf{r}-\mathbf{r}'|).
\end{equation}
With an introduction of the specific PBE semilocal exchange hole model,
 $J^\mathrm{PBE}(s_\sigma(\mathbf{r}), k_{F,\sigma}(\mathbf{r})|\mathbf{r}-\mathbf{r}'|)$,
 and a switch to inter-electron coordinates, $\mathbf{u} = \mathbf{r} - \mathbf{r}'$,
 the semilocal component of the exchange energy is written as the integral
 over a spatially local exchange energy density per electron,
\begin{align} \label{separated_integral}
\epsilon_{\mathrm{x},\sigma}^\mathrm{PBE,SR}(\mathbf{r},\omega) &= 4 \pi \rho_\sigma(\mathbf{r}) \int_0^\infty du \ u  \, \mathrm{erfc}(\omega u) \\
 & \ \ \ \times J^\mathrm{PBE}(s_\sigma(\mathbf{r}), k_{F,\sigma}(\mathbf{r}) u) \notag \\
 E_\mathrm{x}^\mathrm{PBE,SR}(\omega) &=  \sum_{\sigma} \int d\mathbf{r} \rho_\sigma(\mathbf{r})
  \epsilon_{\mathrm{x},\sigma}^\mathrm{PBE,SR}(\mathbf{r},\omega) . \notag
\end{align}
$E_\mathrm{x}^\mathrm{PBE,LR}(\omega)$ has the same form, except with $\mathrm{erfc}(\omega u)$ replaced by $\mathrm{erf}(\omega u)$.
These terms depend on the electron density, $\rho_\sigma (\mathbf{r})$,
 the reduced density gradient, $s_\sigma = |\mathbf{\nabla} \rho_\sigma |/(2 k_{F,\sigma} \rho_\sigma)$, and
 the Fermi wavevector, $k_{F,\sigma} = (6 \pi^2 \rho_\sigma)^{1/3}$.
All equations and results in this paper are in Hartree atomic units.

The $J^\mathrm{PBE}$ function that defines the PBE exchange hole model is non-positive, obeys a normalization constraint,
\begin{equation}\label{Jnorm}
 \int_0^\infty dy \ y^2 J^\mathrm{PBE} (s,y) = -\frac{3 \pi}{4}.
\end{equation}
and reproduces the PBE exchange gradient enhancement factor,
\begin{equation}
 F_\mathrm{x}^\mathrm{PBE}(s) = 1.804 - \frac{0.804}{1 +  0.27302857 s^2},
\end{equation}
 when integrated with the Coulomb interaction,
\begin{equation}\label{Jfit}
 \int_0^\infty dy \ y J^\mathrm{PBE} (s,y) = - \frac{9}{8} F_\mathrm{x}^\mathrm{PBE}(s).
\end{equation}
With these constraints, $J^\mathrm{PBE}$ was fit \cite{PBE_hole} based on properties of the uniform electron gas,
 a principle of minimum information, analytic integrability of Eqs. (\ref{Jnorm}) and (\ref{Jfit}),
 and exact solubility of Eq. (\ref{Jnorm}).
The exchange hole function only appears in calculations
 within integrals containing a Gaussian weight function,
\begin{align} \label{xhole_int}
  &I (s,a)  := - \tfrac{8}{9} \int_0^\infty dy \ y \exp(-a y^2) J^\mathrm{PBE} (s,y) = \\
     & \ \ \ \ c_1 g\left(\frac{f+a}{c_1}\right) + (c_4 + \tfrac{2}{45} f + \tfrac{4}{1215} s^2) \frac{c_2 + f + 5 a}{(c_2 + f + a)^3} \notag \\
   &+   \frac{c_3}{c_2 + f + a} + \frac{\tfrac{16}{15} c_1 (c_2 + f)^3 - \tfrac{8}{15} c_3 (c_2 + f)^2}{(c_2 + f + a)^3}  \notag \\
    &+  \tfrac{16 \sqrt{\pi}}{15}\left[ \tfrac{2}{3} - \sqrt{c_1} \exp \left( \frac{f}{c_1} \right) \mathrm{erfc}\left( \sqrt{\frac{f}{c_1}} \right) \right] \frac{(c_2 + f)^{3.5}}{(c_2 + f + a)^3} \notag \\
 & \ \ \ \ \mathrm{with} \ \ \ \  c_1 = 0.4516064 , \ \ c_2 = 0.57786348 , \notag \\
 & \ \ \ \ \ \ \ \ \ \  \ \ \ \ c_3 = 0.16520372 , \ \ c_4 = 0.0068635965. \notag
\end{align}
Here $g(x)$ contains the exponential integral function, $\mathrm{E}_1(x)$,
\begin{equation}
  g(x) := \ln \left( 1 + \frac{c_2/c_1}{x} \right) - \exp(x) \mathrm{E}_1(x).
\end{equation}
The variable $f$ (denoted $s^2 \mathcal{H}(s)$ in the original paper \cite{PBE_hole})
 is defined implicitly to satisfy Eq. (\ref{Jfit}),
 which now can be written as $I(s,0) = F_\mathrm{x}^\mathrm{PBE}(s)$.
The original version of Eq. (\ref{xhole_int}) was in terms of a different set of five numerical coefficients,
 which we have rearranged into four different coefficients -- $c_1, c_2, c_3, c_4$ -- for compactness and simplicity.

The PBE exchange hole model used by the HSE functional has two mathematical pathologies
 that complicate its use.
First, Eq. (\ref{Jfit}) is satisfied by two values of $f$ for $8.26 \le s \le 11.14$
 and  cannot be satisfied by any value of $f$ for $s > 11.14$.
Second, all partial derivatives of Eq. (\ref{xhole_int}) with $f$ diverge at $f=0$ (when $s=0$)
 while all derivatives of $F_\mathrm{x}^\mathrm{PBE}(s)$ with $s$ are finite.
If $f(s)$ is modeled as $f(s) \propto s^m$ for $s \ll 1$ ($m=4$ was used in the original numerical fit \cite{PBE_hole}),
 the $m^\mathrm{th}$ derivative of Eq. (\ref{xhole_int}) with $s$ will diverge erroneously at $s=0$
 and cannot match the correct response of $F_\mathrm{x}^\mathrm{PBE}(s)$.
The exact $f(s)$ is non-analytic at $s=0$ and approaches zero faster than a power law as $s \rightarrow 0$.

\section{Regularization of the exchange hole}
The primary goals of regularizing the exchange hole model are to satisfy Eq. (\ref{Jfit}) for all $s$
 and to ensure all derivatives of the functional are well-behaved for all $s$.
Secondary goals are simplicity and efficiency.

We regularize the large-$s$ limit without altering the small-$s$ behavior by changing the explicit $s$-dependence in Eq. (\ref{xhole_int})
 from $s$ to
\begin{equation} \label{s_reg}
 \tilde{s}(s) = \left\{ \begin{array}{lr} s & s \le s_1 \\ s  - \exp \left( \ln(s - s_1) - \frac{s_2 - s_1}{s - s_1}\right) & s > s_1 \end{array} \right. ,
\end{equation}
 which smoothly and monotonically limits the range of $s$ from $[0,\infty)$ to $[0,s_2)$.
We make an arbitrary choice of $s_1 := 8$ and $s_2 := 11$,
 which does not alter Eq. (\ref{Jfit}) for small, physical values of $s$ and guarantees that it has a solution for all values of $s$.
This is similar to the previously proposed regularization \cite{HSE09}, but strictly prevents unwanted alterations to the functional for $s \le s_1$.

The regularized form of $f(s)$ is defined to satisfy
\begin{equation} \label{fdef}
 I(\tilde{s}(s),0) = F_\mathrm{x}^\mathrm{PBE}(s).
\end{equation}
The revised $f(s)$ still causes numerical problems with derivatives of $I(s,a)$
 when the non-analytic behavior at $s=0$ is approximated by a rational function.
This issue is resolved by either approximating $f(s)$ with a form that exactly reproduces its non-analytic behavior at $s=0$
 or by numerically evaluating $f(s)$ as a solution of Eq. (\ref{fdef}).
We choose the latter option and apply Newton's method to Eq. (\ref{fdef})
 with $f(s)$ initialized at its previous rational approximation \cite{PBE_hole}.

All the singularities at $f=0$ in Eq. (\ref{xhole_int}) are contained in $g(x)$ at $x=0$.
Its evaluation requires a careful cancellation of logarithmic singularities between the two terms.
An algorithm \cite{E_eval} for the numerical evaluation of $\mathrm{E}_1(x)$ prescribes
 a power series for small $x$ and a continued fraction for large $x$.
We incorporate this result into two series expansions of $g(x)$, for $x \le 2$ as
\begin{align}
 g(x) &= (e^x-1)\ln(x) + \ln(x + c_2/c_1) \\ & \ \ \ + e^x \left( \gamma + \sum_{i=1}^\infty \frac{(-x)^i}{i! i} \right) \notag
\end{align}
 where $\gamma \approx 0.5772156649015329$ is Euler's constant, and
\begin{align}
 g(x) = & \ln \left( 1 + \frac{c_2/c_1}{x} \right) - \frac{1}{b_0-} \frac{a_1}{b_1-} \frac{a_2}{b_2-} \cdots, \\
 & \mathrm{with} \ \ a_i = i^2  \ \ \mathrm{and} \ \ b_i = x + 1 + 2 i, \notag
\end{align}
 for $x > 2$.
To converge $g(x)$ to single precision for all $x$, each expansion requires $\approx 20$ terms.

\section{Numerical integration of the exchange hole}

The difficulty of integrating the exchange hole with the range-separated Coulomb interaction
 originates from the error function in Eq. (\ref{separated_integral}).
The analytic form of the exchange hole was originally chosen by Ernzerhof and Perdew to simplify the integrals
 in Eqs. (\ref{Jnorm}) and (\ref{Jfit}) without considering the difficulty of other integrals.
A simple method for integrating Eq. (\ref{separated_integral}) is to approximate it with something that is analytically integrable.
Specifically, we seek to approximate the complementary error function as a sum of Gaussians,
\begin{equation}
 \mathrm{erfc}(x) \approx \sum_{i=1}^n w_i \exp(-a_i x^2).
\end{equation}
Given such an approximation, we simplify the key integral,
\begin{equation}
 - \tfrac{8}{9} \int_0^\infty dy \ y \mathrm{erfc}(b y) J^\mathrm{PBE} (s,y) \approx \sum_{i=1}^n w_i I(s,a_i b^2).
\end{equation}
Because the exchange hole is non-positive, we construct a simple bound on the error of this approximation,
\begin{align}
  &\tfrac{8}{9} \left| \int_0^\infty dy \ y J^\mathrm{PBE} (s,y) \left(  \mathrm{erfc}(b y) - \sum_{i=1}^n w_i \exp(-a_i b^2 y^2) \right) \right| \notag \\
  &\le  \tfrac{8}{9} \left| \int_0^\infty dy \ y J^\mathrm{PBE} (s,y) \left|  \mathrm{erfc}(b y) - \sum_{i=1}^n w_i \exp(-a_i b^2 y^2) \right| \right| \notag \\
  &\le   \tfrac{8}{9} \varepsilon \left| \int_0^\infty dy \ y J^\mathrm{PBE} (s,y) \right| = \varepsilon F_\mathrm{x}^\mathrm{PBE}(s) 
\end{align}
 with a relative error factor defined as
\begin{equation} \label{min_error}
 \varepsilon := \max_{x \in [0,\infty)} \left| \mathrm{erfc}(x) - \sum_{i=1}^n w_i \exp(-a_i x^2) \right|.
\end{equation}
This analysis naturally limits the numerical errors in the HSE exchange energy
 to be a fraction of the PBE exchange energy.
Also, the error is independent of the scaling of the argument in $\mathrm{erfc}(x)$.

Our numerical integration scheme is optimized by minimizing Eq. (\ref{min_error}) over the choice of $a_i$ and $w_i$,
 which is done once and then tabulated.
This nonlinear minimax optimization is linearized to form
\begin{equation}
 \min_{w_i, v_i} \max_{x\in[0,\infty)} \left| \mathrm{erfc}(x) - \sum_{i=1}^n (w_i - v_i x^2) \exp(-a_i x^2) \right|.
\end{equation}
This is solved using the Remez exchange algorithm, a standard tool in approximation theory \cite{approximation}.
With these solutions, we continuously evolve the exponents as $\delta a_i = v_i / w_i$ until convergence.
This process is not globally convergent and requires a reasonable initial guess for $a_i$,
 which is guided by solutions at smaller values of $n$.
The goal of reducing the error factor $\epsilon$ to below the single precision machine-$\varepsilon$, $1.2 \times 10^{-7}$,
 is achieved for $n=26$ and the coefficients are given in Table \ref{fit_coef}.
There is no need to reduce $\varepsilon$ further because physical modeling errors
 in the HSE functional are orders of magnitude larger than the present numerical errors.

\begin{table}
\centering
\caption{\label{fit_coef} Coefficients that minimize Eq. (\ref{min_error}) for $n = 26$.}
\begin{tabular}{c c c}
 \hline \hline
 $i$ & $w_i$ & $a_i$ \\
 \hline
1 & $2.6444678 \times 10^{-1}$ & $1.0461980 \times 10^{0}$ \\
2 & $2.1518752 \times 10^{-1}$ & $1.4529966 \times 10^{0}$ \\
3 & $1.5602599 \times 10^{-1}$ & $2.4816609 \times 10^{0}$\\
4 & $1.0941077 \times 10^{-1}$ & $4.6668302 \times 10^{0}$\\
5 & $7.6737994 \times 10^{-2}$ & $9.1647121 \times 10^{0}$\\
6 & $5.4113816 \times 10^{-2}$ & $1.8453344 \times 10^{1}$\\
7 & $3.8233937 \times 10^{-2}$ & $3.7911389 \times 10^{1}$\\
8 & $2.6951074 \times 10^{-2}$ & $7.9437595 \times 10^{1}$\\
9 & $1.8891822 \times 10^{-2}$ & $1.6994660 \times 10^{2}$\\
10 & $1.3138701 \times 10^{-2}$ & $3.7184035 \times 10^{2}$\\
11 & $9.0507222 \times 10^{-3}$ & $8.3377708 \times 10^{2}$\\
12 & $6.1667803 \times 10^{-3}$ & $1.9206304 \times 10^{3}$\\
13 & $4.1504553 \times 10^{-3}$ & $4.5580211 \times 10^{3}$\\
14 & $2.7553333 \times 10^{-3}$ & $1.1182009 \times 10^{4}$\\
15 & $1.8013008 \times 10^{-3}$ & $2.8473892 \times 10^{4}$\\
16 & $1.1574260 \times 10^{-3}$ & $7.5635070 \times 10^{4}$\\
17 & $7.2924425 \times 10^{-4}$ & $2.1088326 \times 10^{5}$\\
18 & $4.4921162 \times 10^{-4}$ & $6.2204592 \times 10^{5}$\\
19 & $2.6952465 \times 10^{-4}$ & $1.9611688 \times 10^{6}$\\
20 & $1.5673828 \times 10^{-4}$ & $6.7003372 \times 10^{6}$\\
21 & $8.7754201 \times 10^{-5}$ & $2.5288181 \times 10^{7}$\\
22 & $4.6852295 \times 10^{-5}$ & $1.0845899 \times 10^{8}$\\
23 & $2.3512291 \times 10^{-5}$ & $5.5279640 \times 10^{8}$\\
24 & $1.0830436 \times 10^{-5}$ & $3.6218918 \times 10^{9}$\\
25 & $4.3824327 \times 10^{-6}$ & $3.5946468 \times 10^{10}$\\
26 & $1.4089207 \times 10^{-6}$ & $8.7787680 \times 10^{11}$\\
 \hline \hline
\end{tabular}
\end{table}

With this numerical integration scheme, the HSE semilocal exchange energy density
 from Eq. (\ref{separated_integral}) reduces to
\begin{align}
 \epsilon_{\mathrm{x,\sigma}}^{\mathrm{PBE,SR}}(\mathbf{r},\omega) &= -\frac{3 k_{F,\sigma}(\mathbf{r}) }{4 \pi} \sum_{i=1}^n w_i I \left(\tilde{s}(s_\sigma(\mathbf{r})),\frac{a_i \omega^2}{k_{F,\sigma}(\mathbf{r})^2}\right) \notag \\
 \epsilon_{\mathrm{x,\sigma}}^{\mathrm{PBE,LR}}(\mathbf{r},\omega) &=  -\frac{3 k_{F,\sigma}(\mathbf{r}) }{4 \pi} F_\mathrm{x}^\mathrm{PBE}(s_\sigma(\mathbf{r})) - \epsilon_{\mathrm{x,\sigma}}^{\mathrm{PBE,SR}}(\mathbf{r},\omega).
\end{align}

\section{Accuracy verification for functional and derivatives}

We have incorporated our new implementation of the HSE semilocal exchange functional
 into a development version of the \textsc{libxc} library of DFT functionals \cite{libxc}.
This contribution includes first derivatives, which are mostly straightforward to derive.
Derivatives of $g(x)$ can be related back to its value using the identity $\frac{d}{dx} E_1(x) = \frac{e^{-x}}{-x} $.
For example, its first derivative is
\begin{equation}
 \frac{d}{dx} g(x) = g(x) - \ln\left( 1 + \frac{c_2/c_1}{x}\right) + \frac{1}{c_2/c_1 + x}.
\end{equation}
Derivatives of $f(s)$ can be calculated from its implicit definition in Eq. (\ref{fdef}),
 the first of which is
\begin{equation}
 \frac{d}{d s} f(s) = \frac{\frac{\delta}{\delta s} \left[ F_\mathrm{x}^\mathrm{PBE}(s) - I(\tilde{s}(s),0) \right] }{\frac{\delta}{\delta f} I(\tilde{s}(s),0) }.
\end{equation}

\begin{figure}
\includegraphics[width=90mm]{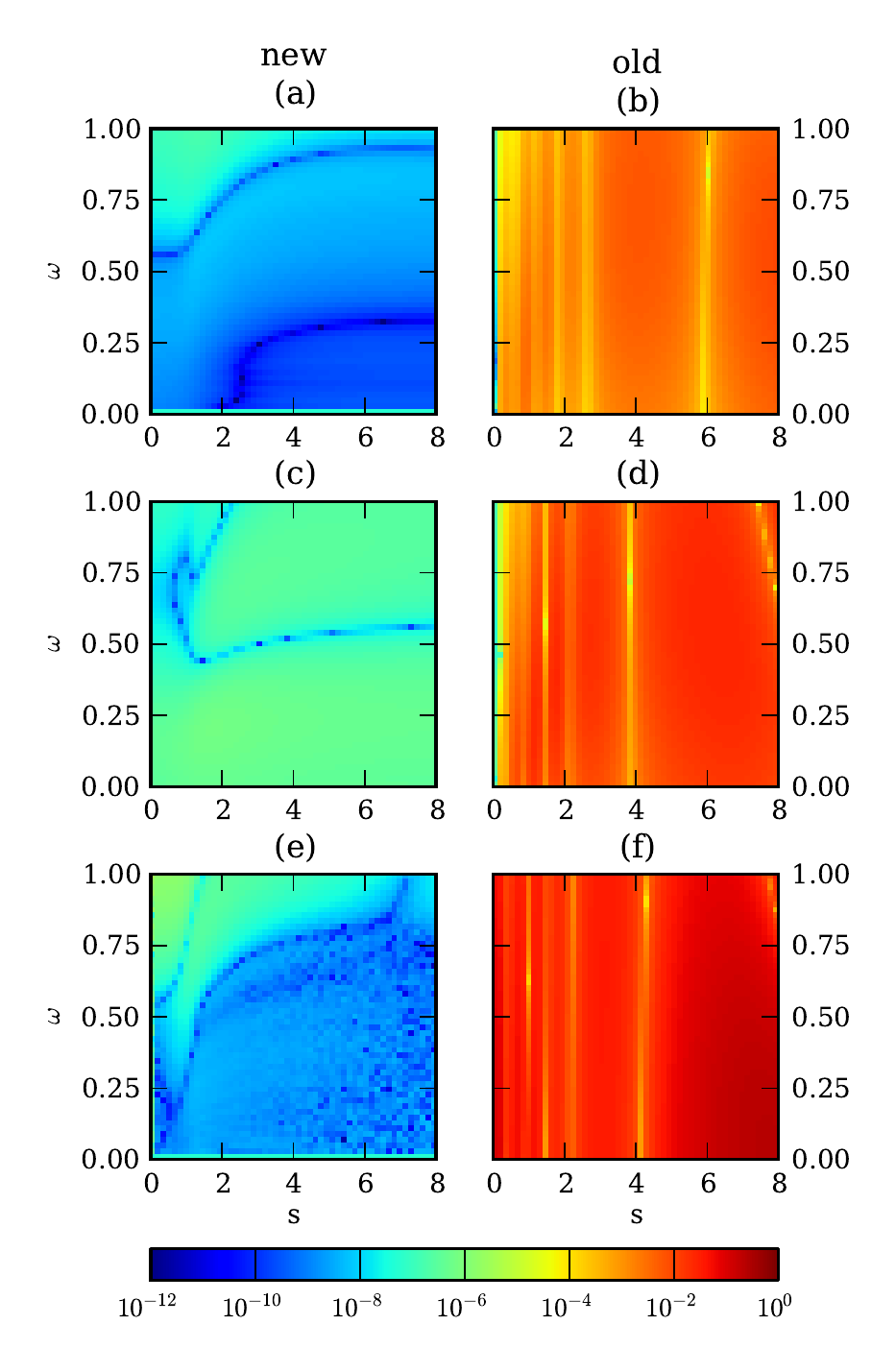}
\caption{\label{accuracy} Accuracy of implementations of the HSE semilocal exchange energy.
The new implementation is shown in the left column (a,c,e), and the old is shown in the right column (b,d,f).
The first row (a,b) shows the energy density per electron, $\epsilon_{\mathrm{x},\sigma}^\mathrm{PBE,SR}(\mathbf{r},\omega)$.
The second row (c,d) shows the density derivative of the energy density per volume, $\frac{d}{d \rho_\sigma(\mathbf{r})} [ \rho_\sigma(\mathbf{r})\epsilon_{\mathrm{x},\sigma}^\mathrm{PBE,SR}(\mathbf{r},\omega)]$.
The third row (e,f) shows the density-gradient derivative of the energy density per volume, $\frac{d}{d |\mathbf{\nabla}\rho_\sigma(\mathbf{r})|} [ \rho_\sigma(\mathbf{r})\epsilon_{\mathrm{x},\sigma}^\mathrm{PBE,SR}(\mathbf{r},\omega)]$.
The relative error between high-accuracy values, $X$, and approximations, $\tilde{X}$, is plotted as $\frac{|X - \tilde{X}| }{ |X| + |\tilde{X}| }$.
We set $\rho$ to a typical material value of $0.01$ and consider a range of parameters: $s \in [0,8]$ and $\omega \in [0,1]$. }
\end{figure}

We compare the new implementation to both a high-accuracy brute-force numerical integration of the exchange hole
 and the second HSE implementation \cite{HSE04}.
While the second version of HSE is not the newest or most accurate,
 it is the most widely used.
It is the only version available in \textsc{vasp} \cite{vasp}, \textsc{espresso} \cite{espresso},
 and the present development version of \textsc{libxc}.
The accuracy of the two implementations is shown in Fig. \ref{accuracy}.
We achieve the target of single precision relative error ($\approx 10^{-7}$) in the functional,
 while the old implementation has relative errors of up to $10^{-2}$.
The errors are amplified by derivatives, but the stringent accuracy requirements of the new implementation
 keep first derivative errors below $\approx 10^{-6}$.
The relative error in the first derivatives gets as large as $0.25$ for the old implementation.

\section{Conclusions}

An important long-term goal of electronic structure research is
 the continued improvement of accuracy within electronic structure simulations.
In this paper, we have demonstrated an instance where numerical errors
 in a popular DFT exchange-correlation functional (HSE) are significant.
We have analyzed these errors, devised an accurate and efficient method to eliminate them,
 and disseminated our new HSE implementation in the open-source \textsc{libxc} library. 
This result suggests that electronic structure research needs to increase
 its awareness of numerical errors and standards of numerical analysis.
Improvements in physical modeling of electron correlation alone cannot lower simulation errors
 below the floor set by numerical errors.

\begin{acknowledgments}
Sandia National Laboratories is a multi-program laboratory managed and  
operated by Sandia Corporation, a wholly owned subsidiary of Lockheed  
Martin Corporation, for the U.S. Department of Energy's National  
Nuclear Security Administration under contract DE-AC04-94AL85000.
\end{acknowledgments}

%
\end{document}